\documentclass[twocolumn,showpacs,nobibnotes]{revtex4}

\usepackage{amsmath}
\usepackage{graphicx}
\usepackage{float}
\usepackage{subfigure}
\usepackage{psfig}
\usepackage{epsfig}

\newcommand{\xgrv}{x_{\text{GRV}}}

\newcommand{\pert}{P}
\newcommand{\nprt}{N\!P}
\newcommand{\WW}{Wegner-Wilson}

\newcommand{\be}{\begin{equation}}
\newcommand{\ee}{\end{equation}}
\newcommand{\bea}{\begin{eqnarray}}
\newcommand{\eea}{\end{eqnarray}}
\newcommand{\benn}{\begin{displaymath}}
\newcommand{\eenn}{\end{displaymath}}
\newcommand{\beann}{\begin{eqnarray*}}
\newcommand{\eeann}{\end{eqnarray*}}
\newcommand{\barray}{\begin{array}}
\newcommand{\earray}{\end{array}}
\newcommand{\inv}{\frac{1}}
\newcommand{\fm}{\mbox{fm}}

\newcommand{\MeV}{\mbox{MeV}}
\newcommand{\GeV}{\mbox{GeV}}

\newcommand{\Identity}{{1\!\rm l}}

\newcommand{\Pc}{{\cal P}}      

\newcommand{\Tr}{\mbox{Tr}}             
\newcommand{\rTr}{\Tr_{r}}        
\newcommand{\nrTr}{\tilde{\Tr}_{r}}             
\newcommand{\tr}{\mbox{tr}}             
\newcommand{\A}{{\!\mbox{\tiny $\cal A$}}}

\newcommand{\befig}{\begin{figure}}
\newcommand{\efig}{\end{figure}}
\newcommand{\betab}{\begin{table}}
\newcommand{\etab}{\end{table}}
\begin{document}

\title{$\mbox{Log}(1/x)$ Gluon Distribution and Structure Functions \\ 
  in the Loop-Loop Correlation Model} 

\author{H.~J.~Pirner$^{1,}$\footnote{E-mail address: pir@tphys.uni-heidelberg.de}} 
\author{A.~I.~Shoshi$^{1,}$\footnote{E-mail address: shoshi@tphys.uni-heidelberg.de}}
\author{G.~Soyez$^{2,}$\footnote{E-mail address: g.soyez@ulg.ac.be}} 
\affiliation{\vspace{0.2cm} $^1$Institut f\"ur Theoretische Physik,
  Universit\"at Heidelberg,\\
  Philosophenweg 19, D-69120 Heidelberg, Germany\\
\vspace{-0.14cm}\\
$^2$Inst. de
  Physique, B\^{a}t. B5, Universit\'{e} de Li\`{e}ge, \\
  Sart-Tilman, B4000 Li\`{e}ge, Belgium}



\begin{abstract}
  
  We consider the interaction of the partonic fluctuation of a scalar
  ``photon'' with an external color field to calculate the leading and
  next-to-leading order gluon distribution of the proton following the
  work done by Dosch-Hebecker-Metz-Pirner~\cite{Dosch:1999yu}. We
  relate these gluon distributions to the short and long distance
  behavior of the cross section of an adjoint dipole scattering off a
  proton.  The leading order result is a constant while the
  next-to-leading order result shows a $\ln(1/x)$ enhancement at small
  $x$. To get numerical results for the gluon distributions at the
  initial scale $Q^2_0=1.8\,\GeV^2$, we compute the adjoint
  dipole-proton cross section in the loop-loop correlation model.
  Quark distributions at the same initial scale are parametrized
  according to Regge theory. We evolve quark and gluon distributions
  to higher $Q^2$ values using the DGLAP equation and compute charm
  and proton structure functions in the small-$x$ region for different
  $Q^2$ values.

  \vspace{.5cm}
 

\end{abstract}


\maketitle
\section{Introduction}\label{sec:intro}

The understanding of deep inelastic scattering (DIS) in the small-$x$
regime remains one of the challenges in quantum chromodynamics (QCD).
Perturbative and nonperturbative physics are important for a complete
picture of the small-$x$ limit of structure functions. In this work,
we combine perturbative and nonperturbative approaches to describe
charm and proton structure functions
(or quark and gluon distributions) 
in the small-$x$ region.

The basic idea is the description of high-energy scattering in QCD by
studying the eikonalized interaction of energetic partons with
external color fields. In ref.~\cite{Dosch:1999yu}, this semiclassical
method has been compared with the parton model to extract the leading
order (LO) and next-to-leading order (NLO) gluon distribution. The
main idea is simple and goes back to Mueller~\cite{Mueller:st}. One
calculates the gluon production cross section for a scalar ``photon'',
which directly couples to the gluon field, in an external color field.
The calculation of one-gluon production leads to the LO gluon
distribution and the one of two-gluon production to the NLO gluon
distribution. The LO result turns out to be a constant,
$xg^{(0)}(x,Q^2) \propto \mbox{const.}$, characterising the averaged
local field strength in the proton. The NLO result shows a logarithmic
increase at small $x$, $xg^{(1)}(x,Q^2) \propto \ln(1/x)$, and is
sensitive to the large distance structure of the proton.
 
In order to make numerical estimates for the gluon distributions, we
relate the LO and NLO gluon distributions to the scattering of a
dipole in the adjoint representation of $SU(3)$ on a proton. Then, we
use the loop-loop correlation model (LLCM)~\cite{Shoshi:2002in} to
compute this adjoint dipole - proton cross section at the center of
mass (c.m.) energy of $\sqrt{s_0} \approx 20\,\GeV$. The two gluons
emerging from the scalar ``photon'' represent the adjoint dipole and a
fundamental quark-diquark dipole models the proton in the LLCM.  The
correlation between the two dipoles given by Wegner-Wilson loops is
evaluated in the Gaussian approximation of gluon field strengths. The
perturbative interactions are obtained from perturbation theory and
the nonperturbative gluon field strength correlator is parametrized in
line with simulations o
in the lattice QCD. The calculation of
scattering cross sections in the loop-loop correlation model has been
quite successful at low energies $\sqrt{s_0}\approx
20\,\GeV$~\cite{Shoshi:2002in,Nachtmann:ua,Nachtmann:ed.kt,Kramer:1990tr,Dosch:1994ym}.

For small virtuality $Q^2$, the perturbative wave function of the two
gluons emerging from the scalar ``photon'' is unrealistically extended
at the endpoints of the longitudinal momentum fraction $\alpha=0,1$.
This problem is similar to the quark-antiquark wave function of the
transverse photon. The phenomenological solution we propose is to give
the gluon a constituent mass $m_G$ of the order of the rho mass.  This
mass modifies the wave function at small transverse gluon momenta
$k_{\!\perp}$ ensuring ``confinement'' for the two gluons, but does not
affect the perturbative part of the two-gluon wave function at high
$k_{\!\perp}$.

We calculate the $x$-dependence of the gluon distribution at a scale
$Q^2_0=1.8\,\GeV^2$ which corresponds to the upper limit where the
nonperturbative input is still credible.  The quark distributions at the
same scale are parametrized in line with Regge theory~\cite{Regge:mz}.
We evolve both distributions to higher values of $Q^2$ using  DGLAP
equations~\cite{Gribov:ri+X}.  With quark and gluon distributions, we
compute charm and proton structure functions at small $x$ for
different $Q^2$ values in good agreement with experimental data.

The outline of the paper is as follows: In section~\ref{sec:semi} we
review the formulas for the calculation of the LO and NLO gluon
distribution following~\cite{Dosch:1999yu}, introduce a gluon mass
into the formalism and rewrite the final expressions for gluon
distributions in terms of adjoint dipole-proton cross sections. We
compute the adjoint dipole-proton cross section in the loop-loop
correlation model in section~\ref{sec:svm}.  In
section~\ref{sec:dglap} we determine the $x$ dependence of the quark
distributions at the initial scale $Q^2_0$ according to Regge theory.
The intial gluon and quark distributions are evolved to higher
virtualities $Q^2$ using the DGLAP equation. In
section~\ref{sec:results} we present the results for the charm and
proton structure function versus Bjorken-$x$ at different $Q^2$
values. Finally, in section~\ref{sec:conclusions}, we summarize our
results.
 
%
\section{The semi-classical gluon distribution}\label{sec:semi}
In this section we review the formulas for the computation of the LO
and NLO gluon distribution~\footnote{Note that the
  expressions "LO" and "NLO" gluon distribution do not refer to the
  order of expansion of the splitting functions in DGLAP evolution. LO
  (resp. NLO) distribution corresponds to one (resp. two) gluon(s) in
  the final state, calculated in the large-$Q^2$ and small-$x$ limit.}
in the semiclassical approach and parton model following the work by
Dosch, Hebecker, Metz and Pirner~\cite{Dosch:1999yu}. In addition, we
introduce a gluon mass in the formalism to take into account the
spatial localization of the two gluons due to confinement. We derive
the modifications due to massive gluons and rewrite the final results
for the gluon distributions in terms of adjoint dipole-proton cross
sections.

\subsection{Gluon distribution at leading order}
In order to extract the gluon distribution, we consider a scalar field
$\chi$ (or scalar ``photon'') coupled directly to the gluon field
through the interaction lagrangian
\[
{\cal L}_I = -\frac{\lambda}{2} \chi \tr\left({\bf F_{\mu\nu}
  F^{\mu\nu}}\right) \ ,
\]
with ${\bf F_{\mu\nu}}= F^a_{\mu\nu} t^a$, the gluon field strength
$F^a_{\mu\nu} = \left [\partial_\mu A^a_\nu - \partial_\nu A^a_\mu - g
  f^{abc}A^b_\mu A^c_\nu \right ]$, the gluon field $A^a$ and the
$SU(N_c)$ group generators in the adjoint representation $t^a$.  In
\begin{figure}[ht]
\centering
\subfigure[]{\includegraphics[scale=0.7]{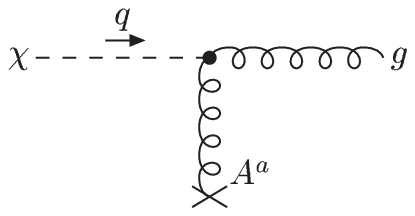}\label{fig:f1a}}\hfill
\subfigure[]{\includegraphics[scale=0.7]{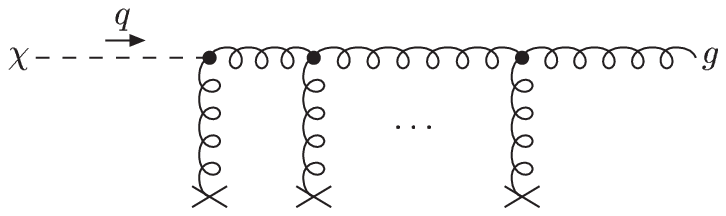}\label{fig:f1b}}
\caption{The process $\chi \to g$ in an external color
  field: \\ (a) one gluon exchange, (b) resummed gluon exchange.}
\label{fig:clLO}
\end{figure}
the high-energy limit, the scattering of a $\chi$-particle off an
external color field $A^a$ as shown in Fig.~\ref{fig:clLO}(a), has the
following amplitude
\[
{\cal T}^a(b_\perp) = \frac{i\lambda q_0}{2C_{\cal A}} \int dx_+ \tr\left[t^a
                       (\epsilon^*_\perp\partial_\perp)
                       A^{\cal A}_{-}(x_+,b_\perp)\right] \ .
\]
Here $b_\perp$ denotes the impact parameter in transverse space, $q$
the four momentum vector of the $\chi$-particle, $x_+ = x_0 + x_3$ the
light-cone variable, $\partial_\perp \equiv \partial / \partial
b_\perp$, $A^{\cal A}=A^b t^b$ the external color field, $C_{\cal A} =
N_c$ the Casimir operator in the adjoint representation, $N_c$ the
number of colors and $\epsilon_\perp$ the polarisation vector of the
outgoing gluon.

Resumming the gluon exchange to all orders as shown in
Fig.~\ref{fig:clLO}(b), the gluon initially created at the vertex
$\chi gg$ aquires a non-abelian eikonal factor on the way through the
external color field
\[
U^{\cal A}_{(\infty,x_+)}(b_\perp) = 
    P\,\exp\left[-\frac{ig}{2}\int_{x_+}^\infty dx_+\,A^{\cal A}_-(x_+, b_\perp) \right].
\]
The path ordering along the way $x_+$ is denoted by $P$.
With
\begin{equation}\label{eq:W}
W^{\cal A}_{b_\perp}(r_\A) = U^{\cal A}(b_\perp - r_\A/2)U^{{\cal
    A}\dagger}(b_\perp+r_\A/2) - \Identity \ ,
\end{equation}
where the eikonal factors $U^{\cal A}(b_\perp- r_\A/2) \equiv U^{\cal
  A}_{\infty,-\infty}(b_\perp- r_\A/2)$ and $U^{{\cal
    A}\dagger}(b_\perp+r_\A/2)$ come from the scattering of two gluons
(moving in opposite directions) with transverse distance $r_\A$ off
the proton, the semiclassical (sc) cross-section for gluon production
at leading order becomes (cf.  ref.~\cite{Dosch:1999yu})
\begin{equation}\label{eq:sc0}
   \sigma_{\text{sc}}^{(0)} = \frac{\lambda^2}{4g^2C_{\cal A}} 
                              \int d^2b_\perp \left| 
                              \left [\partial_{r_\A} W_{b_\perp}^{\cal
                              A}(r_\A)\right ]_{r_\A=0}\right|^2.
\end{equation}
In the following an average over the external color fields underlying the
quantity $W_{b_\perp}^{\cal A}$ is implicitly understood.
  
\begin{figure}[ht]
\centering
\includegraphics[scale=0.7]{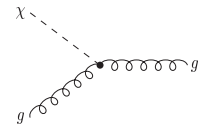}
\caption{The process $\chi g \to g$ in the parton model.}
\label{fig:pmLO}
\end{figure}

On the other hand, the leading-order parton model (pm) cross
section for the partonic process $\chi g \to g$ shown in Fig.~\ref{fig:pmLO} 
with the proton described by the gluon distribution is given by
\begin{equation}\label{eq:pm0}
\sigma_{\text{pm}}^{(0)} = \frac{\pi\lambda}{4} xg^{(0)}(x,Q^2).
\end{equation}
Identifying the semiclassical cross-section~\eqref{eq:sc0} with the
partonic one~\eqref{eq:pm0}, one finds the leading-order gluon
distribution
\begin{equation}\label{eq:gLO}
  xg^{(0)}(x,Q^2) = \frac{1}{2\pi^2\alpha_s} \frac{1}{2C_{\cal A}} \int d^2b_\perp \left| 
                    \left [\partial_{r_\A} W_{b_\perp}^{\cal
                    A}(r_\A)\right ]_{r_\A=0}\right|^2.
\end{equation}
Using the expression for the adjoint dipole (d) - proton (p) cross
section in the semiclassical approach~\cite{Dosch:1999yu,Haas:2000hd}
\begin{equation}
  \sigma^{dp}_{\it sc}(r_\A)   =  -\frac{2}{3} \int d^2b_{\perp}\
                                        \tr \ { W^{\cal
                                        A}_{b_{\perp}}(r_\A)} 
\label{eq:sigdp}
\end{equation} 
and the identity
\begin{equation}
  \int\!\!d^2b_{\!\perp}\!\left|\!\left [\partial_{r_\A} W_{b_\perp}^{\cal
  A}(r_\A)\right ]_{r_\A=0}\right|^2\!\!=\! 
  -\!\left [\partial^2_{r_\A}\!\int\!\!d^2b_{\!\perp} \tr\,{W^{\cal A}_{b_{\perp}}(r_\A)}\!\right ]_{r_\A=0}
\end{equation}
one can rewrite eq.~(\ref{eq:gLO}) in the useful form
\begin{equation} \label{eq:glo1}
   xg^{(0)}(x,Q^2) = \frac{3}{4\pi^2\alpha_s} \frac{1}{2C_{\cal A}}
                     \left [ \partial^2_{r_\A} \sigma^{dp}_{\it sc}(r_\A) 
                     \right]_{r_\A=0} \ .
\end{equation}
This equation shows that the LO gluon distribution depends on the small
distance behavior of the adjoint dipole-proton cross section. To get
numerical results for the gluon distribution, we use in the next
section a model to compute the adjoint dipole-proton cross section.

\subsection{Gluon distribution at next-to-leading order}
\begin{figure}[ht]
\epsfig{figure=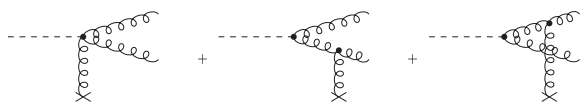, width=8cm}
\caption{Diagrams contributing to the $\chi \to gg$ amplitude in the
  high-energy limit.}
\label{fig:clNLO}
\end{figure}

\begin{figure}[ht]
\epsfig{figure=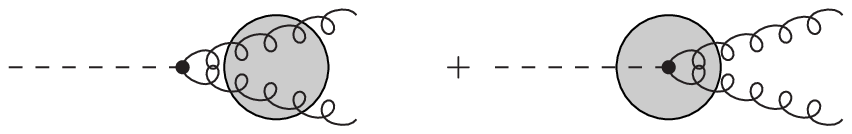, width=8cm}
\caption{Relevant contributions in the NLO semiclassical calculation.}
\label{fig:scNLO}
\end{figure}

For the semiclassical calculation of the gluon distribution at
next-to-leading order, the three diagrams shown in
Fig.~\ref{fig:clNLO} are relevant in the high energy limit. Resumming
the interaction with the external field to all orders, i. e.,
repeating the step leading from Fig.~\ref{fig:clLO}(a) to
Fig.~\ref{fig:clLO}(b), we obtain the result shown in Fig.~\ref{fig:scNLO}.
In the first diagram the incoming $\chi$-particle splits into two
gluons before interacting with the target. The subsequent scattering
of the two gluons off the exteral color field is treated in the
eikonal approximation.  In the second diagram the two fast gluons are
created through a $\chi ggg$ vertex in the space-time region of the
external color field.

One can show~\cite{Dosch:1999yu} that the amplitudes in
Fig.~\ref{fig:scNLO} produce the following NLO contribution to the
cross-section
\begin{eqnarray}\label{eq:scNLO}
\sigma_{\text{sc}}^{(1)}(x,Q^2) 
        & = & \frac{\lambda^2}{32(2\pi)^6}\int 
              \frac{d\alpha}{\alpha(1-\alpha)} \int dk_\perp'^2 \int d^2b_\perp\\
         & &  \left| \int d^2k_\perp \frac{N^2\delta_{ij}+2k_ik_j}{N^2+k_\perp^2+m_G^2} 
              \tilde{W}_{b_\perp}^{\cal A}(k_\perp'-k_\perp)\right|^2, \nonumber
\end{eqnarray}
where $\alpha$ and $1-\alpha$ are the longitudinal momentum carried by
the two gluons, $k_\perp$ and $k_\perp'$ are the transverse momenta of
one of the two gluons before and after the interaction with the
external field, $N^2 = \alpha(1-\alpha) Q^2$ and
$\tilde{W}_{b_\perp}^{\cal A}(k_\perp)$ is the Fourier transform of
$W_{b_\perp}^{\cal A}(r_{\!\mbox{\tiny $\cal A$}})$ given in
eq.~\eqref{eq:W}.

In equation~\eqref{eq:scNLO} we have introduced a gluon mass $m_G$
into the gluon propagator. This mass mimics the confinement of the two
gluons emerging from the $\chi$-particle and modifies their
perturbative wave function in the non-perturbative, small $k_\perp$
region for small virtualities $Q^2$ or momentum fractions $\alpha =
0,1$. Phenomenologically, the gluon mass has a similar effect as a
constituent quark mass in the perturbative quark-antiquark wave
function of the transverse photon~\cite{Dosch:1997nw}. We use for the
gluon mass $m_G = 770\,\MeV$, i.e., approximately half of the glueball
mass.

\begin{figure}[ht]
\epsfig{figure=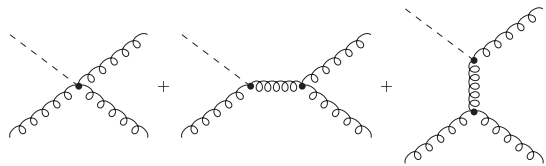, width=8cm}
\caption{Diagrams contributing to the parton model at NLO.}
\label{fig:pmNLO}
\end{figure}

To compute the NLO contribution to the gluon distribution in the
parton model, we have to calculate the diagrams shown in Fig.~\ref{fig:pmNLO}. 
With
$z=Q^2/(Q^2+s)=x/y$, where $y$ is the fraction of the target momentum
carried by the struck gluon, the parton model cross-section at NLO
reads
\be \label{eq:sigtot_pm}
  \sigma_{\text{pm}}(x, Q^2) = \int_x^1 \frac{dz}{z}\left[\sigma^{(0)}_d \delta(1-x) 
                               + \hat{\sigma}^{(1)}(x,Q^2)\right]
                               yg_b(y), 
\ee
with $\sigma^{(0)}_d = (\pi \lambda_d^2)/4$ and the $d =
(4+\varepsilon)$ dimensional coupling $\lambda_d = \lambda
\mu^{-\varepsilon / 2}$. Since we are interested in the high energy limit, we
extract only the leading term in $\ln(1/x)$ from the
expression in~(\ref{eq:sigtot_pm}). We regularize the bare
gluon distribution $g_b(x)$ in the $\overline{\rm{MS}}$ scheme 
\begin{eqnarray*}
g_b(x) & = & g(x,\mu^2) \\
       & - & \frac{\alpha_s}{2\pi} \int_x^1 \frac{dz}{z} P_{gg}(z) 
       \left[ \frac{2}{\varepsilon} + \gamma_E - \ln(4\pi)\right] g(y, \mu^2),
\end{eqnarray*}
with $P_{gg}$ denoting the gluon-gluon splitting function.
The parton model cross-section at NLO finally becomes
\begin{eqnarray}\label{eq:pmNLO}
  \sigma_{\text{pm}} & = & \sigma^{(0)}_{d=4} x \int_x^1 \frac{dz}{z} g(y, \mu^2)\\
                       && \left\{\delta(1-z) + \frac{\alpha_s}{2\pi}\left[ P_{gg}(z) 
                          \ln\left(\frac{Q^2}{\mu^2}\right) +
                          C_g^{\overline{\rm{MS}}}(z)\right] \right\},\nonumber
\end{eqnarray}
where
\be
C_g^{\overline{\rm{MS}}}(z) =
     P_{gg}(z) \ln \left (\frac{1-z}{z}\right ) - \frac{11 C_{\cal A}}{6 z (1-z)} \ .\nonumber
\ee
Expanding the full gluon distribution up to the order $\ln(1/x)$  
\[
xg(x,\mu^2) = xg^{(0)}(x, \mu^2) + xg^{(1)}(x, \mu^2), 
\]
and identifying~\eqref{eq:scNLO} with~\eqref{eq:pmNLO}, we obtain 
\begin{eqnarray}\label{eq:e1}
&&   xg^{(1)}(x, \mu^2) = \frac{1}{4(2\pi)^7} \int
                \frac{d\alpha}{\alpha(1-\alpha)} 
                \int dk_\perp'^2 \int d^2b_\perp \nonumber \\ 
             & & \times \left| \int d^2k_\perp
                \frac{N^2\delta_{ij}+2k_ik_j}{N^2+k_\perp^2+m_G^2} 
                \tilde{W}_{b_\perp}^{\cal A}(k_\perp'-k_\perp)\right|^2 \nonumber\\
             & &-\frac{\alpha_s}{2\pi} \int_x^1 dz \left[ P_{gg}(z) 
                   \ln\left(\frac{Q^2}{\mu^2}\right) + 
                  C_g^{\overline{\rm{MS}}}(z)\right]yg^{(0)}(y, \mu^2)
                \ , \nonumber \\
\end{eqnarray}
where $g^{(0)}(x, \mu^2)$ is given by~\eqref{eq:gLO}. When evaluating
equation~\eqref{eq:e1}, we keep only $\ln(1/x)$ terms, i.e., we use
\begin{eqnarray*}
P_{gg}(z) & \approx & \frac{2C_{\cal A}}{z}, \\
C_g^{\overline{\rm{MS}}}(z) & \approx & 
\frac{2C_{\cal A}}{z}\left[ \ln\left(\frac{1}{z}\right) - \frac{11}{12} \right].
\end{eqnarray*}
Using the variable $z=Q^2/(Q^2+M^2)$, with $M^2 =
k_\perp'^2/(\alpha(1-\alpha))$, and introducing a scale $\kappa^2$
such that $\Lambda_{QCD}^2 \ll \kappa^2 \ll Q^2$, the first
contribution in~\eqref{eq:e1} can be split into a hard and a soft
component~\cite{Dosch:1999yu,Haas:2000hd}
\begin{eqnarray}\label{eq:e2}
\!\!\!\!\!\!\!\!\!\!\!&& \frac{1}{4(2\pi)^7} \int
                \frac{d\alpha}{\alpha(1-\alpha)} 
                \int dk_\perp'^2 \int d^2b_\perp \nonumber \\ 
\!\!\!\!\!\!\!\!\!\!\!&& \times \left| \int d^2k_\perp
                \frac{N^2\delta_{ij}+2k_ik_j}{N^2+k_\perp^2+m_G^2} 
                \tilde{W}_{b_\perp}^{\cal
                A}(k_\perp'-k_\perp)\right|^2 
                \nonumber\\
\!\!\!\!\!\!\!\!\!\!\!&=& \lefteqn{\frac{1}{4\pi^3} \int_x^1 \frac{dz}{z} 
                \ln\left(\frac{Q^2}{z\kappa^2}\right) \int d^2b_\perp \left| 
                \left[\partial_{r_\A} W_{b_\perp}^{\cal A}(r_\A)\right]_{r_\A=0}\right|^2}\nonumber\\
\!\!\!\!\!\!\!\!\!\!\!&& + \ \frac{2}{\pi} \int_x^1 \frac{dz}{z} 
                   \int_0^{\kappa^2}dk_\perp'^2\,f(k_\perp'^2) \ ,
\end{eqnarray}
where 
\begin{eqnarray*}
   f(k_\perp'^2) & = & \int\frac{d^2r_\A}{(2\pi)^2r_\A^2} 
                       \int\frac{d^2r^{\prime}_\A}{(2\pi)^2r^{\prime 2}_\A}  \int d^2b_\perp \\
                  & &  \tr\left[W_{b_\perp}^{\cal A}(r_\A)
                       W_{b_\perp}^{{\cal A}\dagger}(r^{\prime}_\A)\right] 
                       e^{ik_\perp'(r_\A-r^{\prime}_\A)} H(r_\A,r^{\prime}_\A),\\
               H & = & \frac{(r_\A \cdot r^{\prime}_\A)^2}{r_\A^2
                       r^{\prime 2}_\A} 
                       \left[{\hat a}^2 K_0({\hat a})\right] 
                       \left[{\hat b}^2 K_0({\hat b})\right]\\
                 & + & \frac{1}{2}\left[\frac{2(r_\A \cdot
                       r^{\prime}_\A)^2}{r_\A^2 r^{\prime 2}_\A}-1\right] 
                       \left\{\left[{\hat a} K_1({\hat a})\right]
                       \left[{\hat b} K_1({\hat b})\right]\right . \\
                   &&  \phantom{\left[\frac{2(r_\A \cdot r^{\prime}_\A)^2}
                      {r_\A^2 r^{\prime 2}_\A}-1\right]} + 
                       \left[{\hat a} K_1({\hat a})\right] 
                       \left[{\hat b}^2 K_0({\hat b})\right] \\
                   &&  \phantom{\left[\frac{2(r_\A \cdot r^{\prime}_\A)^2}
                       {r_\A^2 r_2^2}-1\right]}\left. 
                       + \left[{\hat a}^2 K_0({\hat a})\right] 
                       \left[{\hat b} K_1({\hat b})\right]\right\},
\end{eqnarray*}
with ${\hat a} = m_G r_\A$ and ${\hat b} = m_G r^{\prime}_\A$.  Note that the
gluon mass does not influence the hard part which is calculated at
leading twist. The lower bound of the $z$ integration, $x \leq z$, is
a kinematical limit ensuring that the invariant mass of the two produced
gluons cannot be larger than the total center-of-mass energy
available~\cite{Dosch:1999yu,Haas:2000hd}.

Inserting~\eqref{eq:e2} in~\eqref{eq:e1}, the $\ln^2(1/x)$ terms 
from the semiclassical and the parton calculations cancel, so that  
\begin{eqnarray*}
   xg^{(1)}(x,\mu^2) & = & \ln\left(\frac{1}{x}\right) \left\{ \frac{2}{\pi} 
                           \int_0^{\kappa^2}dk_\perp'^2\,f(k_\perp'^2)\right.\\
                      & &  \phantom{\ln 1 } +
                           \left.\frac{\alpha_s}{\pi}C_{\cal A}\left[\ln\left(\frac{\kappa^2}{\mu^2}\right)+
                           \frac{11}{12}\right] xg^{(0)}\right\}.
\end{eqnarray*}
For $\kappa^2 = \mu^2\exp(11/12)$, the second term also drops out and
one obtains 
\begin{equation}\label{eq:g1}
     xg^{(1)}(x,\mu^2) = \frac{2}{\pi}  \ln\left(\frac{1}{x}\right) 
                \int_0^{e^{11/12}\mu^2}dk_\perp'^2\,f(k_\perp'^2).
\end{equation}
After the integration over $k_\perp'^2$ and $r'_{\cal A}$, the NLO correction
to the gluon distribution becomes~\cite{Dosch:1999yu,Haas:2000hd}
\begin{eqnarray}\label{eq:gnlo}
  xg^{(1)}(x,\mu^2) & = & \frac{1}{2\pi^3}\ln\left(\frac{1}{x}\right)
                          \int_{r_0^2(\mu^2)}^\infty 
                          \frac{dr_\A^2}{r_\A^2}\,m_G^2\,F(m_G r_\A)\nonumber\\
                     & &  \times \int d^2b_\perp\tr\left[W_{b_\perp}^{\cal A}
                          (r_\A)W_{b_\perp}^{{\cal A}\dagger}(r_\A)\right]
\end{eqnarray}
where
\begin{eqnarray*}
   r_0^2(\mu^2) & = & \frac{4e^{\frac{1}{12}-2\gamma_E}}{\mu^2}, \\
   F(z) & = & K_1^2(z) + zK_0(z)K_1(z) + \frac{z^2}{2} K_0^2(z) \\
        & = & \frac{z^2}{2}K_2^2(z) - zK_1(z)K_2(z) + K_1^2(z).
\end{eqnarray*}
In the large-$N_c$ limit~\cite{Buchmuller:1998jv}, one obtains the
relation  
\begin{equation}
\int d^2b_\perp \tr \left[W_{b_\perp}^{\cal A}
             (r_\A)W_{b_\perp}^{{\cal A}\dagger}(r_\A)\right] =
- 2 \int\!d^2b_{\!\perp}\ \tr \ { W^{\cal A}_{b_{\perp}}(r_\A)} \nonumber
\end{equation}
which allows us to rewrite the NLO correction in terms 
of the dipole-proton cross section~(\ref{eq:sigdp})
\begin{eqnarray}\label{eq:gnlo1}
  xg^{(1)}(x,\mu^2) \!&\!=\!&\! \ln\left(\frac{1}{x}\right) 
                          \frac{3 m_G^2}{2\pi^2}
                          \int_{r_0^2(\mu^2)}^\infty 
                          \frac{dr_\A^2}{r_\A^2} F(m_G r_\A)
                          \ \sigma^{dp}_{\it sc}(r_\A) . \nonumber
                          \\ 
\end{eqnarray}
The NLO gluon distribution depends on the large distance behavior of
the adjoint dipole-proton cross section. For small $r_{\cal A}$ the
function $F(m_G r_{\cal A})$ reduces to $1/(m_G r_{\cal A})^2$, and
eq.~(15) shows qualitative agreement with eqs.~(49)-(50)
of Mueller~\cite{Mueller:st}. To obtain numerical results for the LO and NLO
gluon distribution, we calculate the dipole-proton cross section at
some low c.m. energy $\sqrt{s_0}$ within the loop-loop correlation
model in the following section.

\section{Dipole-Proton Cross Section from the Loop-Loop Correlation
         Model}\label{sec:svm}
In this section we use the loop-loop correlation model
(LLCM)~\cite{Shoshi:2002in} to compute the cross section of an
adjoint dipole scattering off a proton, $\sigma^{dp}_{\it
sc}(r_\A, s_0)$, at the c.m. energy $\sqrt{s_0} = 20\,\GeV$.  
Then, we insert the resulting
dipole-proton cross section in eqs.~(\ref{eq:glo1}) and
(\ref{eq:gnlo1}) to calculate the LO and NLO gluon
distributions.

The loop-loop correlation model is based on the functional integral
approach to high-energy scattering in the eikonal
approximation~\cite{Nachtmann:ua,Nachtmann:ed.kt,Kramer:1990tr,Dosch:1994ym}.
Its central elements are gauge-invariant \WW \ loops. With a
phenomenological energy dependence, we have shown that the LLCM allows
a unified description of high-energy hadron-hadron, photon-hadron, and
photon-photon reactions~\cite{Shoshi:2002in,Shoshi:2002fw,Shoshi:2002fq} and of
static properties of hadrons~\cite{Shoshi:2002rd,Steffen:2003rd} in agreement with
experimental and lattice data. In this work, we do not need a
phenomenological parametrization for the energy dependence since this
is generated by the formalism outlined in the sections before.

In the framework of LLCM the adjoint dipole - proton cross section
reads~\cite{Shoshi:2002in,Shoshi:2002fw}
\begin{eqnarray}
  \sigma^{dp}_{LLCM}(r_{\!\mbox{\tiny $\cal A$}}, s_0) &=& 2\int d^2b_{\!\perp}\!\int
      \frac{d\phi_{\!\mbox{\tiny $\cal A$}}}{2 \pi} \int d^2 r_{\!\mbox{\tiny $\cal F$}} dz_q |\psi_p(z_q,{\vec r}_{\!\mbox{\tiny $\cal F$}})|^2
      \nonumber \\
     && \times \left(1 -  S^{\cal A\,\cal F}(s_0,{\vec
      b}_{\!\perp},{\vec r}_{\!\mbox{\tiny $\cal A$}},z_q,{\vec r}_{\!\mbox{\tiny $\cal F$}})\right) \ .
\label{siddp_llcm}
\end{eqnarray}
Here the correlation of two light-like Wegner-Wilson loops
\be
        S^{\cal A\,\cal F}(s_0,{\vec b}_{\!\perp},{\vec r}_{\!\mbox{\tiny $\cal A$}},z_q,{\vec r}_{\!\mbox{\tiny $\cal F$}})
        = \frac{\Big\langle {\cal W}^{\cal A}[C_g] {\cal W}^{\cal F}[C_q] \Big\rangle_G}
               {\Big\langle {\cal W}^{\cal A}[C_g]\Big\rangle_G
               \Big\langle {\cal W}^{\cal F}[C_q] \Big\rangle_G} \ , 
\label{Eq_loop_loop_correlation_function}
\ee 
describes the elastic scattering of a dipole in the fundamental ($\cal
F$) with a dipole in the adjoint ($\cal A$) representation of
$SU(N_c)$. In the present case where a $\chi$-particle scatters off a
proton, the adjoint color-dipole is given by the two gluons in the
color-singlet state emerging from the $\chi$-particle and the
fundamental color-dipole is given in a simplified picture by a quark
and diquark in the proton~\footnote{The proton has in fact a
  three-quark structure and is described by three
  dipoles~\cite{Dosch:1994ym,Kharzeev:1996sq} instead of a
  quark-diquark dipole as done here. However the quark-diquark
  description simplifies enormously the model and gives similar
  results as the three-dipole picture once the model parameters are
  readjusted~\cite{Dosch:1994ym}.}.  Each color-dipole is represented
by a light-like Wegner-Wilson loop~\cite{Wilson:1974sk+X}
\be
        {\cal W}^r[C] = 
        \nrTr\,\Pc
        \exp\!\left[
        -i g \oint_{\scriptsize C} dz^{\mu}\,A_{\mu}^a(z)\,t_r^a 
        \right]      
        \ , \nonumber
\ee
where the subscript $r$ indicates the representation of $SU(N_c)$,
$\nrTr = \rTr(\cdots)/\Tr \Identity_r$ is the normalized trace in the
corresponding color-space with unit element $\Identity_r$, and
$A_{\mu}(z) = A_{\mu}^a(z) t_r^a$ represents the gluon field with the
$SU(N_c)$ group generators in the corresponding representation,
$t_r^a$, that demand the path ordering indicated by $\Pc$ on the
closed path $C$ in space-time. Physically, the {\WW} loop represents
the phase factor acquired by a color-charge in the $SU(N_c)$
representation $r$ along the light-like trajectory $C$ in the gluon
background field.

The color-dipoles have transverse size and orientation ${\vec
  r}_{\!\mbox{\tiny $\cal A$,$\cal F$}}$.
The longitudinal momentum fraction of the dipole carried by the
quark (gluon) is $z_q$ ($z_g$). The impact parameter between 
the dipoles is~\cite{Dosch:1997ss}
\be
        {\vec b}_{\!\perp} 
        \,=\, {\vec r}_{g} + (1-z_g) {\vec r}_{\!\mbox{\tiny $\cal A$}} 
            - {\vec r}_{q} - (1-z_q) {\vec r}_{\!\mbox{\tiny $\cal F$}} 
        \,=\, {\vec r}_{\!\mbox{\tiny $\cal A$}\,cm} - 
              {\vec r}_{\!\mbox{\tiny $\cal F$}\,cm} 
        \ , \nonumber 
\ee
where ${\vec r}_{q}$, ${\vec r}_{qq}$, ${\vec r}_{g}$, and ${\vec
  r}_{\bar{g}}$ are the transverse positions of the quark, diquark,
gluon and the gluon moving in the opposite direction, respectively.
With ${\vec r}_{\!\mbox{\tiny $\cal F$}} = {\vec r}_{qq} - {\vec
  r}_{q}$ and ${\vec r}_{\!\mbox{\tiny $\cal A$}} = {\vec r}_{\bar{g}}
- {\vec r}_{g}$, the center of light-cone momenta
of the two dipoles are given by ${\vec r}_{\!\mbox{\tiny $\cal A$}\,cm} =
z_{g} {\vec r}_{g} + (1-z_{g}){\vec r}_{\bar{g}}$ and ${\vec
  r}_{\!\mbox{\tiny $\cal F$}\,cm} = z_{q} {\vec r}_{q} +
(1-z_{q}){\vec r}_{qq}$.
\befig[ht]
  \begin{center}
        \epsfig{file=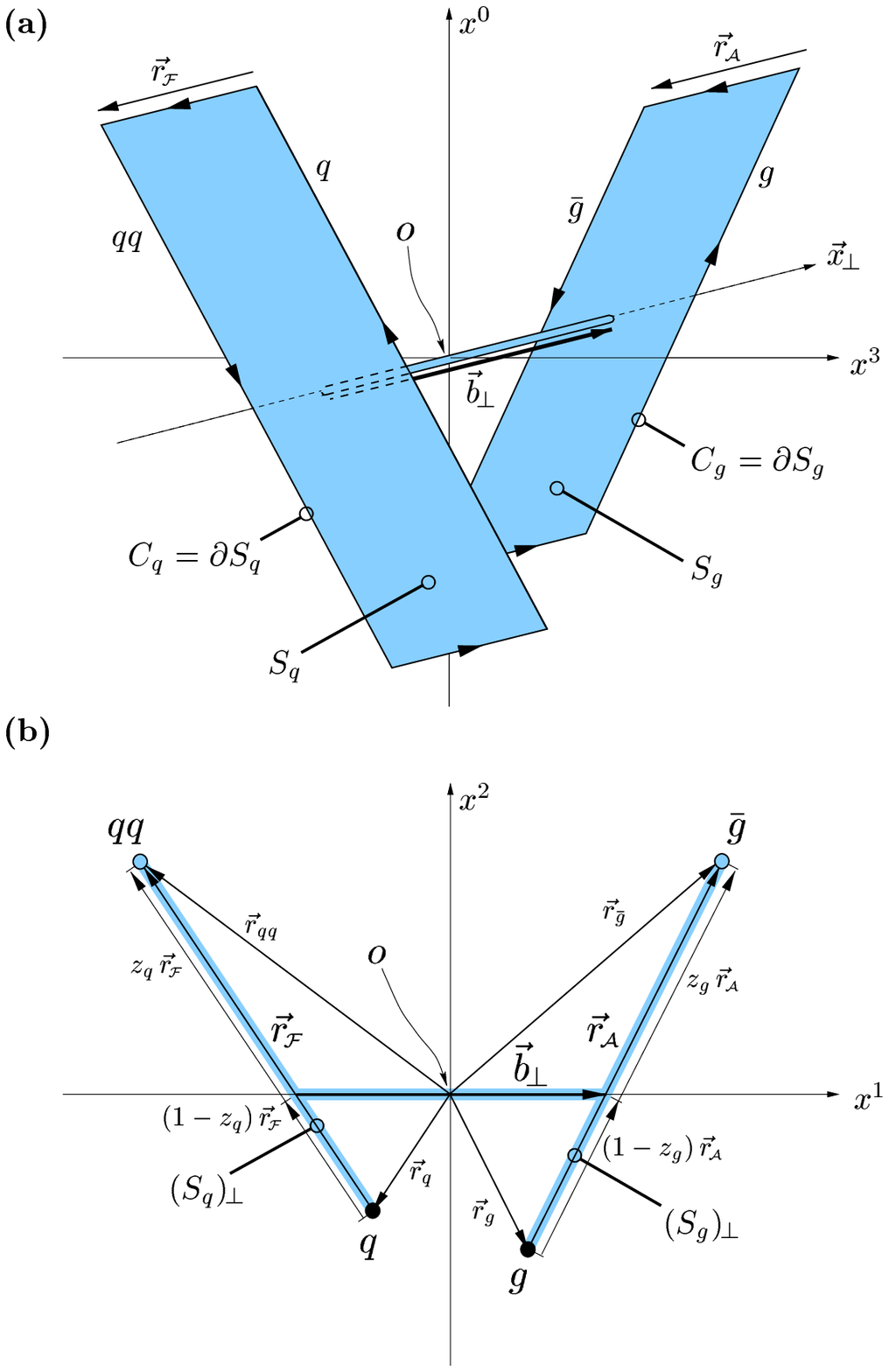,width=8.cm}
  \end{center}
\caption{\small High-energy scattering of a fundamental dipole with 
  an adjoint dipole in the eikonal approximation represented by
  Wegner-Wilson loops: (a) space-time and (b) transverse arrangement
  of the Wegner-Wilson loops. The shaded areas represent the strings
  extending from the quark (gluon) to the diquark (antigluon) path in each color dipole.
  The thin tube allows to compare the field strengths in surface $S_q$
  with the field strengths in surface $S_g$. The impact parameter
  $\vec{b}_{\perp}$ connects the centers of light-cone momenta of the
  dipoles.}
\label{Fig_loop_loop_scattering_surfaces}
\efig
 Figure~\ref{Fig_loop_loop_scattering_surfaces} illustrates the (a)
space-time and (b) transverse arrangement of the dipoles. 

The QCD vacuum expectation value $\langle \ldots \rangle_G$ in the
loop-loop correlation function
(\ref{Eq_loop_loop_correlation_function}) represents functional
integrals~\cite{Nachtmann:ed.kt} over gluon field configurations: the
functional integration over the fermion fields has already been
carried out as indicated by the subscript $G$.  The model we use for
the QCD vacuum describes only gluon dynamics and, thus, implies the
quenched approximation that does not allow string breaking through
dynamical quark-antiquark production.

The ${\vec r}_{\!\mbox{\tiny $\cal F$}}$ and $z_q$ distribution of the fundamental color-dipole in the
proton is given by the proton wave function $\psi_p$.  We use
for the proton wave function the phenomenological Gaussian
Wirbel-Stech-Bauer ansatz~\cite{Wirbel:1985ji}
\be
        \psi_p(z_q,\vec{r}_{\!\mbox{\tiny $\cal F$}}) 
        = \sqrt{\frac{z_q(1-z_q)}{2 \pi S_p^2 N_p}}\, 
        e^{-(z_q-\inv{2})^2 / (4 \Delta z_p^2)}\,  
        e^{-|\vec{r}_{\!\mbox{\tiny $\cal F$}}|^2 / (4 S_p^2)} 
        \ . \nonumber
\ee
The constant $N_p$ is fixed by the normalization
of the wave function to unity, the extension parameter
$S_p$ is approximately given by the electromagnetic radius of the
proton and the width $\Delta z_p =
w/(\sqrt{2}\,m_p)$~\cite{Wirbel:1985ji} is determined by the proton
mass $m_p$ and the value $w = 0.35 - 0.5\,\GeV$ extracted from
experimental data. We adopt the values $\Delta z_p= 0.3$ and $S_p =
0.86\,\fm$ which have allowed a good description of many high-energy
scattering data in our previous work~\cite{Shoshi:2002in}.

The computation of the loop-loop correlation for one dipole in the
fundamental and the other one in the adjoint representation of
$SU(N_C)$ can be found in detail
in~\cite{Shoshi:2002fw,Shoshi:2002rd,Steffen:2003rd}. The main steeps
of this computation are the transformation of the line integrals in
the \WW \ loops into surface integrals with the non-Abelian Stokes
theorem, a matrix cumulant expansion, and the Gaussian approximation
of the functional integrals in the gluon field strengths. These steeps
lead to the following result
\bea
\!\!\!\!\!\!\!\!\!\!\!\!\!
\frac{\Big\langle {\cal W}^{\cal A}[C_g] {\cal W}^{\cal F}[C_q] \Big\rangle_G}
               {\Big\langle {\cal W}^{\cal A}[C_g]\Big\rangle_G
               \Big\langle {\cal W}^{\cal F}[C_q] \Big\rangle_G}
         &=& \,
        \,\inv{N_c^2\!-\!1}\,\exp\!\Big[i\,\frac{N_c}{2}\,\chi\Big]
        \nonumber \\
        &+& \frac{N_c\!+\!2}{2(N_c\!+\!1)}\exp\!\Big[\!-i\,\inv{2}\chi\Big]
        \nonumber \\
        &+&
    \frac{N_c\!-\!2}{2(N_c\!-\!1)}\exp\!\Big[i\,\inv{2}\chi\Big] 
\label{loop_loop_correlations}
\eea
which, for $N_c=3$, corresponds to the well-known $SU(3)$ decomposition
\be
        3\,\otimes\,8 
        = 3\,\oplus\,15\,\oplus\,6
        \ . \nonumber
\ee
The function ${\chi}:= \chi^{\pert} + \chi^{\nprt}_{nc} + \chi^{\nprt}_{c}$
has the following form~\cite{Shoshi:2002in}:
\bea 
    \chi^{\pert} \!\!&=&\!\!  \left[
         g^2\!\left(\vec{r}_{g}-\vec{r}_{q}\right) iD^{\prime\,(2)}_{\pert}
         \left(\vec{r}_{g}-\vec{r}_{q}\right) \right.
         \nonumber \\
         &+&\,g^2\!\left(\vec{r}_{\bar{g}}-\vec{r}_{qq}\right)
         iD^{\prime\,(2)}_{\pert} 
         \left(\vec{r}_{\bar{g}}-\vec{r}_{qq}\right)
         \nonumber \\
         &-&\,g^2\!\left(\vec{r}_{g}-\vec{r}_{qq}\right)
         iD^{\prime\,(2)}_{\pert} \left(\vec{r}_{g}-\vec{r}_{qq}\right)
         \nonumber \\
         &-& \left. g^2\!\left(\vec{r}_{\bar{g}}-\vec{r}_{q}\right)
         iD^{\prime\,(2)}_{\pert} \left(\vec{r}_{\bar{g}}-\vec{r}_{q}\right)
         \right] ,
\label{Eq_chi_PGE}\\ \vspace{2cm} \nonumber \\
\!\!
        \chi_{nc}^{\nprt} \!\!&=&\!\!\! 
        \frac{\pi^2 G_2 (1\!-\!\kappa)}{3(N_c^2-1)} 
        \!\left[ 
        iD^{\prime\,(2)}_1\!
        \left(\vec{r}_{g}-\vec{r}_{q}\right)\! 
        +iD^{\prime\,(2)}_1\!
        \left(\vec{r}_{\bar{g}}-\vec{r}_{qq}\right)
        \right.
        \nonumber \\
        &&\!\!\!\!\!\!\!\!\!  
        \hphantom{\chi_{nc}^{\nprt}=\frac{\pi^2}{3}}
        \left.
        -\,iD^{\prime\,(2)}_1\!
        \left(\vec{r}_{g}-\vec{r}_{qq}\right)
        -iD^{\prime\,(2)}_1\!
        \left(\vec{r}_{\bar{g}}-\vec{r}_{q}\right)
        \right] , \nonumber \\
\label{Eq_chi_MSV_non-confining}\\ \vspace{2cm} \nonumber \\
      \chi_{c}^{\nprt} \!\!&=&\!\! 
        \frac{\pi^2 G_2 \kappa}{3(N_c^2-1)}
        \left(\vec{r}_{\!\mbox{\tiny $\cal A$}}\cdot\vec{r}_{\!\mbox{\tiny $\cal F$}}\right) 
\label{Eq_chi_MSV_confining} \\
        &&\!\!\times \int_0^1 \! dv_{\!\mbox{\tiny $\cal A$}} \int_0^1 \! dv_{\!\mbox{\tiny $\cal F$}} \, 
        iD^{(2)}\left(\vec{r}_{g} + v_{\!\mbox{\tiny $\cal A$}}\vec{r}_{\!\mbox{\tiny $\cal A$}} 
        - \vec{r}_{q} - v_{\!\mbox{\tiny $\cal F$}}\vec{r}_{\!\mbox{\tiny $\cal F$}}\right) \ , \nonumber
\eea
with the perturbative ($iD^{\prime\,(2)}_{\pert}$) and nonperturbative
($iD^{\prime\,(2)}_1$ and $iD^{(2)}(\vec{z}_{\!\perp})$)
correlation functions in transverse space 
\bea
        iD^{\prime\,(2)}_{\pert}
        (\vec{z}_{\!\perp})
        \!\!&=&\!\! \inv{2\pi} K_0\left(m_G |\vec{z}_{\!\perp}|\right)
        , 
\label{Eq_F2[i_massive_D_pge_prime]} \\ \vspace{0cm} \nonumber \\
  iD^{\prime\,(2)}_1(\vec{z}_{\!\perp})
   \!\!&=&\!\!  \pi  a^4 \!\left[3\!+\!3(|\vec{z}_{\!\perp}|/a)\!+\! 
                 (|\vec{z}_{\!\perp}|/a)^2 \right]
                 \exp\!\left( -|\vec{z}_{\!\perp}|/a\right)
                 , \nonumber \\
\label{Eq_F2[i_D_non-confining_prime]} \\ \vspace{0cm} \nonumber \\
        iD^{(2)}(\vec{z}_{\!\perp})
        \!\!&=&\!\!  2 \pi \, a^2 
        \left[1+(|\vec{z}_{\!\perp}|/a)\right] 
        \exp\!\left( -|\vec{z}_{\!\perp}|/a\right) \ .
\label{Eq_F2[i_D_confining]}
\eea
We have introduced in the perturbative component $\chi^{\pert}$ the
same effective gluon mass $m_G = 0.77\,\GeV$ as before to limit the range
of the perturbative interaction in the infrared region and a parameter
$M^2 = 1.04\,\GeV^2$ which freezes the running coupling in the
quenched approximation at the value $g^2(\vec{z}_{\!\perp})/(4 \pi) =
\alpha_s(k_{\!\perp}^2=0) = 0.4$~\cite{Shoshi:2002in},
\be
        g^2(\vec{z}_{\!\perp})
        = \frac{48 \pi^2}
        {(33-2 N_f) 
        \ln\left[
                (|\vec{z}_{\!\perp}|^{-2} + M^2)/\Lambda_{QCD}^2
        \right]} \ . 
\label{Eq_g2(z_perp)}
\ee
In Eq.~(\ref{Eq_F2[i_massive_D_pge_prime]}) $K_0$ denotes the $0^{th}$
modified Bessel function (McDonald function).  The non-perturbative
correlators~(\ref{Eq_chi_MSV_non-confining})
and~(\ref{Eq_chi_MSV_confining}) involve the gluon condensate $G_2 :=
\langle \frac{g^2}{4\pi^2} F^a_{\mu\nu}(0) F^a_{\mu\nu}(0) \rangle =
0.074\,\GeV^4$, the parameter $\kappa= 0.74$ that determines the
relative weight of the two different components and the correlation
length $a = 0.302\,\fm$ that enters through the non-perturbative
correlation functions $D$ and $D_1$.

The component $\chi^{\pert}$ describes the perturbative interaction of
the quark and diquark of the dipole in the proton with the two gluons
of the adjoint dipole emerging from the $\chi$-particle as evident
from~(\ref{Eq_chi_PGE}) and
Fig.~\ref{Fig_loop_loop_scattering_surfaces}b. The component
$\chi_{nc}^{\nprt}$ has the same structure as $\chi^{\pert}$ and gives
the non-perturbative interaction between the quarks and gluons of the
two dipoles. The component $\chi_{c}^{\nprt}$ shows a different
structure: the integrations over $v_{\!\mbox{\tiny $\cal A$}}$ and
$v_{\!\mbox{\tiny $\cal F$}}$ sum non-perturbative interactions
between the strings (flux tubes) confining the quark and diquark or
the two gluons in the dipoles as visualized in
Fig.~\ref{Fig_loop_loop_scattering_surfaces}b. As shown in
~\cite{DelDebbio:1994zn,Rueter:1995cn,Shoshi:2002rd}, the
$\chi_{c}^{\nprt}$ component leads to color-confinement due to a flux
tube formation between a static quark-antiquark pair. Manifestations
of confinement in high-energy scattering have been analysed in~\cite{Shoshi:2002fq}.

Since the squared proton wave function
$|\psi_p(z_q,\vec{r}_{\!\mbox{\tiny $\cal F$}})|^2$ is invariant and
the $\chi$-function changes sign under the replacement
$(\vec{r}_{\!\mbox{\tiny $\cal F$}} \rightarrow
-\vec{r}_{\!\mbox{\tiny $\cal F$}}, z_q \rightarrow 1-z_q)$
\be \chi(\vec{b}_{\!\perp},z_g,\vec{r}_{\!\mbox{\tiny $\cal A$}},1-z_q,-\vec{r}_{\!\mbox{\tiny $\cal F$}}) = -
\chi(\vec{b}_{\!\perp},z_g,\vec{r}_{\!\mbox{\tiny $\cal A$}},z_q,\vec{r}_{\!\mbox{\tiny $\cal F$}}) \ , \nonumber
\label{Eq_odd_eikonal_function}
\ee
only the real part of the exponentials in
eq.~(\ref{loop_loop_correlations}) survives in the integration over
${\vec r}_{\!\mbox{\tiny $\cal F$}}$ and $z_q$ so that one obtains
\bea 
\!\!\!\!\!\!\!\!\!\!\!\!\!
       \frac{\Big\langle {\cal W}^{\cal A}[C_g] {\cal W}^{\cal F}[C_q] \Big\rangle_G}
               {\Big\langle {\cal W}^{\cal A}[C_g]\Big\rangle_G
               \Big\langle {\cal W}^{\cal F}[C_q] \Big\rangle_G}
   &=&\, \,\inv{N_c^2\!-\!1}\,\cos\Big[\frac{N_c}{2}\,\chi\Big]
   \nonumber \\
   &+& \frac{N_c\!+\!2}{2(N_c\!+\!1)}\cos\Big[\inv{2}\chi\Big]
   \nonumber \\
   &+& \frac{N_c\!-\!2}{2(N_c\!-\!1)}\cos\Big[\inv{2}\chi\Big] \ .
\eea
The above expression describes multiple gluonic interactions between
two dipoles since $(\chi^{\pert})^2$ represents the perturbatively
well-known two-gluon exchange and $(\chi^{\nprt})^2$ the
non-perturbative two-point interaction in the dipole-dipole
scattering~\cite{Shoshi:2002fq}. The higher order terms in the
expansion of the cosine functions ensure the $S$-matrix unitarity
condition which becomes important at very high c.m.
energies~\cite{Shoshi:2002in,Shoshi:2002fw,Shoshi:2002mt+X}.

The adjoint dipole-proton cross section obtained with the above
ingredients at the c.m. energy $\sqrt{s_0} \approx 20\,\GeV$ shows
color-transparency for small dipole sizes,
\be
\sigma^{dp}_{LLCM}(s_0,r_{\!\mbox{\tiny $\cal A$}}) \approx 9.6\, r_{\!\mbox{\tiny $\cal A$}}^2 
\label{ctransparency}
\ee
and linear confining behavior at large dipole sizes,
\be
\sigma^{dp}_{LLCM}(s_0,r_{\!\mbox{\tiny $\cal A$}}) \propto |\vec{r}_{\!\mbox{\tiny $\cal A$}}|  \ .
\ee
Inserting eqs.~(\ref{ctransparency}) and~(\ref{Eq_g2(z_perp)})
into~(\ref{eq:glo1}), one obtaines for the LO gluon
distribution~(\ref{eq:glo1}) at virtuality $Q^2_0 = 1.8\,\GeV^2$
\begin{equation} \label{eq:glo11}
   xg^{(0)}(x,Q^2_0) = \frac{3}{4\pi^2\alpha_s(Q^2_0)}
   \frac{1}{2C_{\cal A}}\,2 \cdot 9.6 = 0.81\ .
\end{equation}
With our result for $\sigma^{dp}_{LLCM}(s_0,r_{\!\mbox{\tiny $\cal
    A$}})$, the NLO gluon distribution~(\ref{eq:gnlo1}) at the same
virtuality reads
\be \label{eq:gnlo11}
  xg^{(1)}(x,Q^2_0) = 0.89 \ \ln\left(\frac{1}{x}\right) \ . 
\ee
%

\section{DGLAP evolution at high $Q^2$}\label{sec:dglap}
In this section we give the gluon and quark distributions at an
initial scale $Q_0^2$. Their evolution to higher values of $Q^2$ is
obtained by the DGLAP equation
\begin{eqnarray}\label{eq:DGLAPfull}
\lefteqn{ Q^2\partial_{Q^2}
\begin{pmatrix}q_i(x,Q^2)\\\bar q_i(x,Q^2)\\g(x,Q^2)\end{pmatrix}}\\
& = &\frac{\alpha_s}{2\pi} \int_x^1 \frac{d\xi}{\xi}
\left.\begin{pmatrix}
 P_{q_iq_j} & 0 & P_{q_ig}\\
 0 & P_{q_iq_j} & P_{q_ig}\\
 P_{gq} & P_{gq} & P_{gg}
\end{pmatrix}\right|_{\frac{x}{\xi}}
\begin{pmatrix} q_j(\xi,Q^2)\\\bar q_j(\xi,Q^2)\\ g(\xi, Q^2)
\end{pmatrix}
 \ , \nonumber
\end{eqnarray}
with the splitting functions $P_{xy}$ being at leading order $Q^2$
independent. We use the resulting gluon and quark distributions to
compute the charm and proton structure function at different $x$ and
$Q^2$ values.

The gluon distribution computed in the previous section reads 
\begin{equation}\label{eq:scg}
   xg(x, Q_0^2) = A\left[ 1 + B \ln(1/x)\right] \ ,
\end{equation}
with $A = 0.81$ and $B=1.1$ for $Q_0^2 = 1.8\,\GeV^2$. Our result,
however, is only expected to be valid at low $x$ values. For large $x$
values, $x > \xgrv = 0.15$, we use the Gluck-Reya-Vogt (GRV) gluon
distribution~\cite{Gluck:1998xa}. To match to this gluon
distribution at $x = \xgrv$, we introduce a scale $x_0$ in our gluon
distribution
\begin{equation}\label{eq:g2}
  G(x,Q_0^2) = xg(x, Q_0^2) = A\left[ 1 + B \ln(x_0/x)\right] 
\end{equation}
which takes into account the neglected constant term in the NLO
calculation where only the leading $\ln(1/x)$ terms have been kept.
For $Q_0^2=1.8\,\GeV^2$ and $\xgrv=0.15$, we obtain $x_0 = 0.1454$. 

To calculate the proton structure function $F_2^p(x,Q^2)$, 
\[
F_2^p(x) = x \sum_{\text{flavours}} e_q^2\left[q(x) + \bar{q}(x)\right],
\]
with the DGLAP evolution~\eqref{eq:DGLAPfull}, we need quark and
gluon distributions, since they are coupled to each other. For the
computation of $F_2^p(x,Q^2)$, however, only two linear combinations
of quark distributions are required
\begin{eqnarray*}
   T   & = & x(u^++c^++t^+)-x(d^++s^++b^+),\\
\Sigma & = & x(u^++c^++t^+)+x(d^++s^++b^+),
\end{eqnarray*}
since
\[
  F_2^p = \frac{5\Sigma + 3T}{18} \ ,
\]
with $q^+ = q+\bar{q}$ and $q = u, d, s, c, t, b$.  Performing linear
combinations in~\eqref{eq:DGLAPfull}, we can directly check that $T$,
$\Sigma$ and $G$ evolve according to the following DGLAP equations
\begin{eqnarray*}
Q^2 \partial_{Q^2}\!\!\!&&\!\!\! T(x,Q^2) 
    =  \frac{\alpha_s}{2\pi} \int_x^1 \frac{xd\xi}{\xi^2} 
         P_{qq}\left(\frac{x}{\xi}\right)
         T(\xi,Q^2), \\
Q^2 \partial_{Q^2}\!\!\!&&\!\!\!\begin{pmatrix} \Sigma(x,Q^2) \\ G(x,Q^2)
   \end{pmatrix} 
   \nonumber \\
   &&= \frac{\alpha_s}{2\pi} \int_x^1 \frac{xd\xi}{\xi^2}
         \left. 
         \begin{pmatrix} P_{qq} & 2n_fP_{qg}\\
                         P_{gq} & P_{gg}
         \end{pmatrix}\right|_{\frac{x}{\xi}}
         \begin{pmatrix} \Sigma(\xi,Q^2) \\ G(\xi,Q^2) \end{pmatrix},
\end{eqnarray*}
which means that $T$ evolves alone, while $\Sigma$ is coupled with $G$.

We determine the $x$-dependence of the $T$ and $\Sigma$ distribution
at the scale $Q^2_0$ following the ideas from~\cite{Soyez:2002nm}.
First, the $T$, $\Sigma$ and $G$ distributions are required to have a
common singularity structure in the complex $j$-plane according to
Regge theory. Since the initial gluon distribution~\eqref{eq:g2} has a
double pole in $j=1$, consequently, also the $T$ and $\Sigma$
distribution are required to have a double-pole pomeron term.
Secondly, we add a reggeon contribution (coming from the exchange of
meson trajectories $a_0$ and $f$) to the quark distribution. In the
gluon distribution we negelect this term since the reggeon is expected
to be constituted of quarks. Moreover, we expect that the
pomeron, having vacuum quantum numbers, does not distinguish between
quark flavours, i. e., the pomeron decouples from the $T$
distribution. In contrast, the $\Sigma$ distribution contains a
pomeron and a reggeon component. The initial distributions, thus, read
\begin{eqnarray}
  T(x,Q_0^2) & = & \tau x^{\alpha_0} (1-x)^\sigma,\\
  \Sigma(x, Q_0^2) & = & \left[ a \ln(1/x) + b + 
                         dx^{\alpha_0} \right](1-x)^\kappa  \nonumber
\end{eqnarray}
with the reggeon intercept $\alpha_0=0.4$ and the powers $\sigma=3$
and $\kappa=2$ of $(1-x)$ taking into account daughter trajectories in
Regge theory. For $x>\xgrv$, we again rely on the GRV distributions
since our $T$ and $\Sigma$ distributions are only valid at small $x$.
The parameters $\tau$ and $b$ are fixed to ensure continuity
between our distributions and GRV's ones at $x=\xgrv$. Finally, we are
left with 2 parameters, $a$ and $d$, which we determine by fitting the
$F_2^p(x,Q^2)$ experimental data.

\section{Results}\label{sec:results}
\begin{figure}
\includegraphics{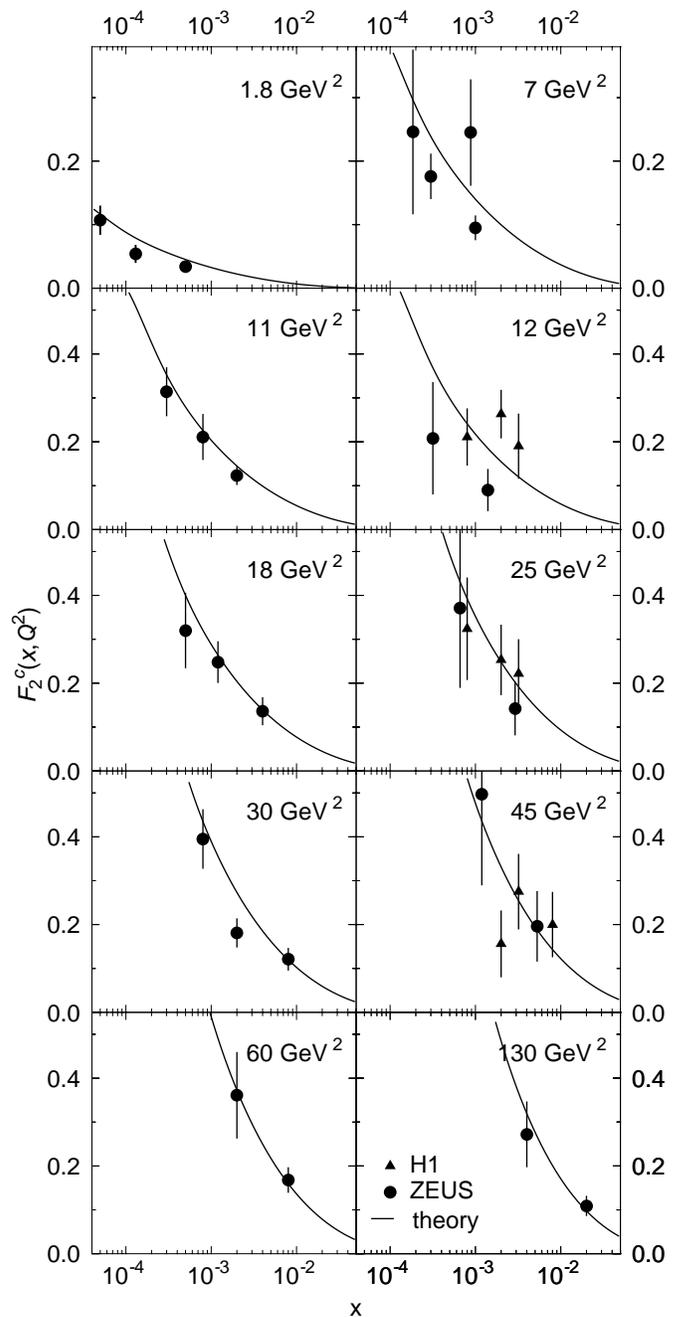}
\caption{The charm structure function, $F_2^c(x,Q^2)$, as a function
  of the Bjorken-variable $x$ at different virtualities $Q^2$.}
\label{fig:fc}
\end{figure}
\begin{figure*}
\includegraphics{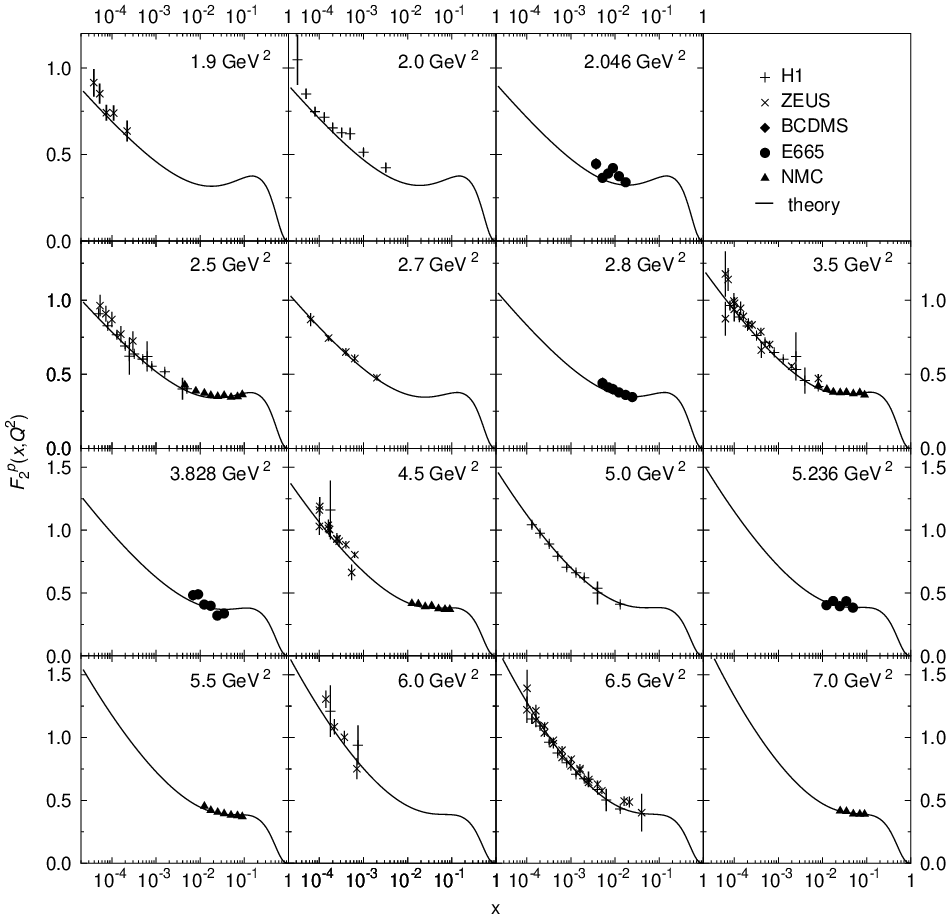}
\caption{The proton structure function, $F_2^p(x,Q^2)$, as a function
  of the Bjorken-variable $x$ at low virtualities $Q^2$.}
\label{fig:f2-low}
\end{figure*}
\begin{figure*}
\includegraphics{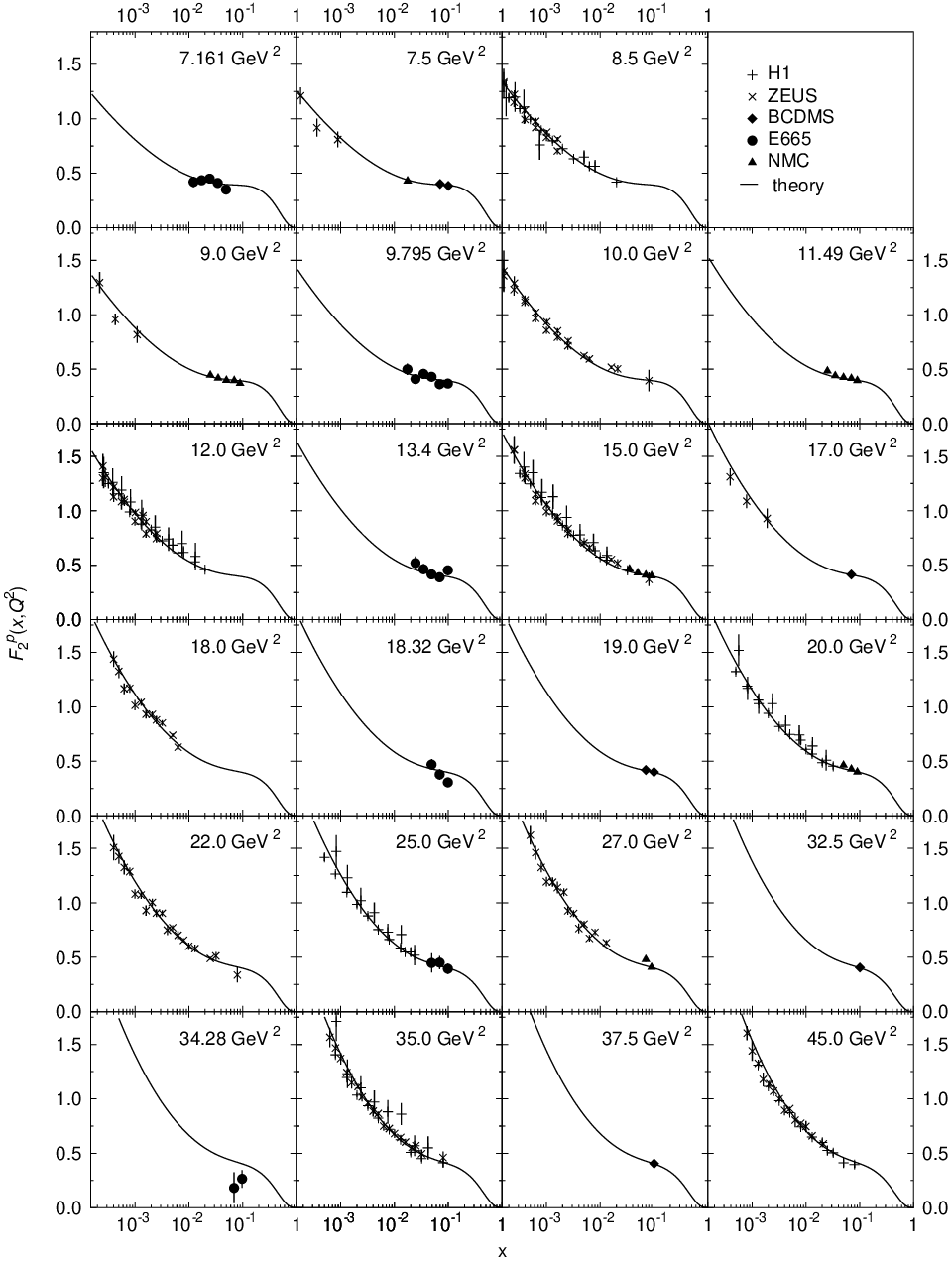}
\caption{The proton structure function, $F_2^p(x,Q^2)$, as a function
  of the Bjorken-variable $x$ at middle-range virtualities $Q^2$.}
\label{fig:f2-mid}
\end{figure*}
\begin{figure*}
\includegraphics{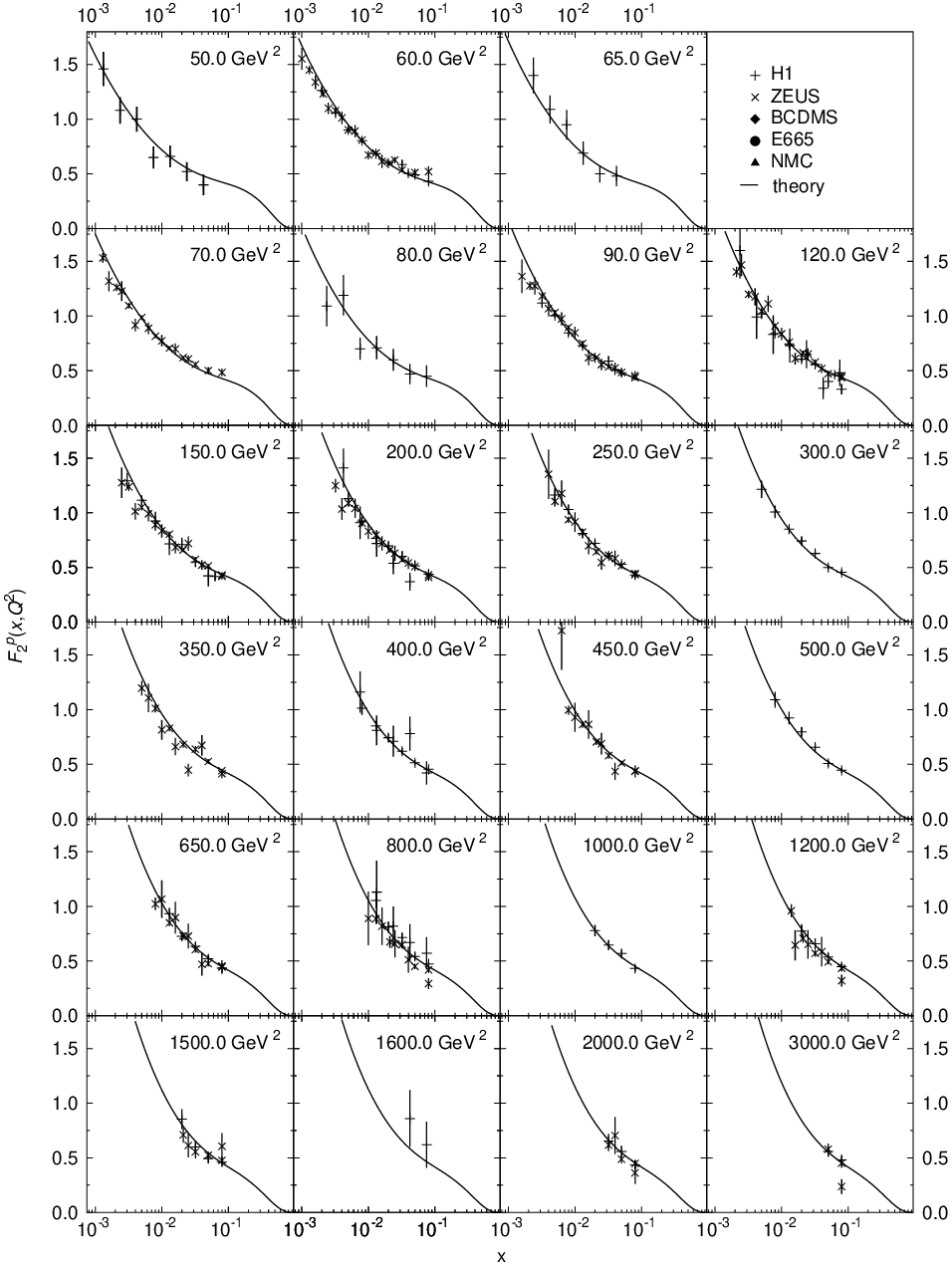}
\caption{The proton structure function, $F_2^p(x,Q^2)$, as a function
  of the Bjorken-variable $x$ at high virtualities $Q^2$.}
\label{fig:f2-high}
\end{figure*}

With the previous $T$, $\Sigma$ and $G$ distributions at $Q^2_0
=1.8\,\GeV^2$ as an initial condition in the DGLAP equation, we have
fitted the $F_2^p$ experimental data within the domain
\begin{eqnarray}
   Q^2 & \ge & Q_0^2 = 1.8\,\GeV^2 \ ,\nonumber\\
     x & \le & \xgrv = 0.15 \ ,\\ 
     \cos(\theta_t) & = & \frac{\sqrt{Q^2}}{2\,x\,m_p} \ge
     \frac{49\,\GeV^2}{2m_p^2} \ .\nonumber
\end{eqnarray}
Here $m_p$ denotes the proton mass, $\theta_t$ is the scattering angle
in the $t$-channel and $\cos(\theta_t)$ has been extended to the whole
complex plane. The last condition of $\cos(\theta_t)$ taken from
\cite{Cudell:2002ej} ensures that Regge theory is applicable. 
We have used 922 experimental points coming from
H1~\cite{Adloff:1999ah,Adloff:2000qj,Adloff:2000qk},
ZEUS~\cite{Breitweg:1999ad,Chekanov:2001qu},
BCDMS~\cite{Benvenuti:1989rh}, E665~\cite{Adams:1996gu} and
NMC~\cite{NMC} to adjust our parameters within the given region. 
We have obtained the following values
\begin{eqnarray*}
      a & = & 0.45933 \pm 0.00425 \ ,\\
      b & = & -1.9598 \qquad \text{(fixed by continuity)} \ ,\\
      d & = & 6.0408 \pm 0.0456 \ ,\\
      \tau & = & 0.55628 \qquad \text{(fixed by continuity)} \ .\\
\end{eqnarray*}

In Figs.~\ref{fig:f2-low}, \ref{fig:f2-mid} and \ref{fig:f2-high} we
show the results for the proton structure function, $F_2^p(x,Q^2)$, as
a function of the Bjorken-variable $x$ between the virtualities
$Q^2=1.8\,\GeV^2$ and $Q^2=3000\,\GeV^2$. Our results are in good
agreement with experimental data for all $x$ and $Q^2$ values.

To test our gluon distribution, we compute the charm structure
function which depends on the gluon distribution as
follows~\cite{book:roberts}
\begin{eqnarray*}
  F_2^c(x,Q^2) & = & 2e_c^2\frac{\alpha_s(Q^2+4m_c^2)}{2\pi}\\
               &   & \int_{ax}^1 d\xi\,g(\xi, Q^2+4m_c^2)f(x/\xi, Q^2)
               \ ,
\end{eqnarray*}
with
\begin{eqnarray*}
  f(x,Q^2) & = & v\left\lbrack(4-\mu)x^2(1-x)-\frac{x}{2}\right\rbrack \\
           & + & L \left\lbrack \frac{x}{2}- x^2(1-x)+\mu
           x^2(1-3x)-\mu^2z^3\right\rbrack \ ,\\
       \mu & = & \frac{2m_c^2}{Q^2} \ , \\
         v & = & \sqrt{1-\frac{2x\mu}{1-x}} \ ,\\
         L & = & \log\left(\frac{1+v}{1-v}\right) \ ,\\
         a & = & 1+2\mu \ .
\end{eqnarray*}
We have adopted a value of 1.25 GeV for the charm quark mass. The
predictions for $F_2^c(x,Q^2)$ obtained from our model are presented
together with the experimental HERA
data~\cite{Adloff:1996xq,Breitweg:1997mj,Breitweg:1999ad} in
Fig.~\ref{fig:fc}. The good agreement of our predictions for
$F_2^c(x,Q^2)$ with experimental data clearly shows that we obtain a
resonable gluon distribution.

\section{Conclusions}\label{sec:conclusions}
We have considered the interaction of a scalar ``photon'', which
directly couples to the gluons, with an external color field to
extract the leading and next-to-leading order gluon distribution. We
have closely followed the previous work by Dosch, Hebecker, Metz and
Pirner~\cite{Dosch:1999yu}, in which the semiclassical approach (where
the partonic fluctuations of the ``photon'' interact with the proton
in the eikonal approximation) has been compared with the parton model
to get the gluon distribution of the proton. The leading order result
is a constant while the next-to-leading order result has a $\ln(1/x)$
rise at small $x$.

We have been able to relate the leading and next-to-leading order
gluon distribution to the short and long distance behavior of the
cross section of a dipole in the adjoint representation of $SU(3)$
scattering off a proton, respectively. In addition, a gluon mass has
been introduced to take into account nonperturbative effects in the
small-$k_{\!\perp}$ region of the perturbatively derived wave function
of the two gluons emerging from the scalar ``photon'', in analogy to
the constituent quark mass in the quark-antiquark wave function. We
have computed the adjoint dipole-proton cross section in the loop-loop
correlation model~\cite{Shoshi:2002in} to obtain numerical results
for the gluon distributions at the initial scale $Q^2_0=1.8\,\GeV^2$.

Quark distributions at the same initial scale have been parametrized
in line with Regge theory and the ideas leading to the gluon
distribution.  We have used DGLAP equations to evolve quark and gluon
distributions to higher $Q^2$ values. The charm and proton structure
functions computed in the small-$x$ region are in good agreement with
experimental HERA data over a large range of $Q^2$ values.

\begin{acknowledgments}
  We would like to thank Arthur Hebecker and Stephane Munier for
  stimulating discussions. This research is partially funded by the
  INTAS project ``Non-Perturbative QCD'' and the European TMR Contract
  HPRN-CT-2000-00130. G.~S. is supported by the National Fund for
  Scientific Research (FNRS), Belgium.
\end{acknowledgments}


\end{document}